\documentclass[12pt]{article}
\usepackage[utf8]{inputenc}
\usepackage{amscd,amsfonts,amstext,amsmath,amssymb,amsthm,bm}
\usepackage[numbers]{natbib}
\usepackage{mathtools}
\usepackage{subcaption}
\usepackage{framed}
\usepackage{url}
\usepackage{color}

\newcommand{\esp}{\mathbb{E}}
\newcommand{\var}{\mathbb{V}}
\newcommand{\varf}{{\bf V}} 

\newcommand{\dkappa}{\dot{\kappa}}
\newcommand{\ddkappa}{\ddot{\kappa}}
\newcommand{\realn}{\mathbb{R}}

\newcommand{\nat}{\mathbb{N}}

\newcommand{\Beta}{\mathcal{B}}
\newcommand{\floor}[1]{\left\lfloor #1 \right\rfloor}

\newcommand{\highlight}[1]{ #1}
\newcommand{\correctedComa}[1]{#1}

\newtheorem{definition}{Definition}[section]

\title{The Unifed Distribution}

\author{Oscar Alberto Quijano Xacur\footnote{Correspondence: oscar.quijano@use.startmail.com}
   \\
Concordia University, Montreal, Canada}

\begin{document}
\maketitle

\begin{abstract}
  We introduce a new distribution with support on (0,1) called
  unifed. It can be used as the response distribution for a GLM and it
  is suitable for data aggregation. We make a comparison to the beta
  regression. A link to an R package for working with the unifed is
  provided.
\end{abstract}


\section*{Introduction}

We introduce the unifed distribution. It is a continuous distribution
with support on the interval (0,1). It can be characterized as the
only exponential dispersion family containing the uniform
distribution. This makes it suitable to be used as the response
variable of a Generalized Linear Model (GLM).

An R (see \cite{r-language} and \cite{unifed-rpackage}) package has
been developed to work with this distribution. It is called
\verb|unifed| and contains functions for the density, distribution,
quantiles and random generator. It also contains a family that can be
used within the \verb|glm| function of R. Additionally, the package
provides \highlight{Stan} \cite{rstan} code for performing Bayesian
analysis with the unifed including a function for fitting Bayesian
unifed GLMs. Information about the package and how to install it can
be found at \url{https://gitlab.com/oquijano/unifed}.

This is not the only model for performing regression on the unit
interval. The beta regression (see \cite{beta-regression}) has existed
for a while and it provides more flexible shapes than the unifed
GLM. One appealing property of the unifed GLM is that it is suitable
for data reduction while the beta regression is not. This is discussed
in section 4.2.

This paper is divided \highlight{into} 4 sections. In Section 1 we
review the definition and properties of exponential dispersion
families and GLMs. Section 2 defines the unifed distribution. In
Section 3 we illustrate an application to an auto insurance claims
example. Section 4 reviews the beta regression and underlines it's
differences with the unifed GLM.

\section{Exponential Dispersion Families and GLMs}
\label{sec:edf-glms}

A reproductive Exponential Dispersion Family (EDF) is a set of
distributions whose densities are given by
\begin{equation}
  \label{eq:edf-density}
  f(y|\theta,\phi) = a(y,\phi)\exp\left(\frac{1}{\phi} \left\{y\theta
      - \kappa(\theta) \right\}\right),\qquad \theta\in\Theta, \phi\in\Phi\/.
\end{equation}
$\theta\/$ and $\Theta\/$ are called the canonical parameter and
canonical space, respectively and $\phi\/$ is known as the dispersion
parameter. For $\theta\in\mbox{int} \left( \Theta\right)\/$ (here int
stands for interior),
\begin{equation}
  \esp[Y] = \dot{\kappa}\left(\theta\right)\qquad \mbox{and}\qquad  \var[Y]=\phi\ddot{\kappa}\left(\theta\right)\/,
  \label{eq:kappa-derivatives}
\end{equation}
where $\dot{\kappa}=\kappa'$ and
$\ddot{\kappa}=\dot{\kappa}'$. \eqref{eq:kappa-derivatives} allows to
relate the mean and the variance and the mean of any EDF. This motivates the following
definitions (see \cite{Jorgensen-book} or \cite{jorgensen-deviance}).

\begin{definition}
  Given an exponential dispersion family, the mean domain of the
  family is defined as
  \begin{equation*}
    \Omega = \left\{\mu = \dot{\kappa}\left(\theta\right) : \theta \in \textrm{int}\left(\Theta\right) \right\}\/.
  \end{equation*}
\end{definition}

\begin{definition}
  The \emph{variance function} of \highlight{an} EDF is defined as
  $\varf: \Omega \rightarrow [0,\infty)$ with
  \begin{equation*}
    \varf(\mu) = (\ddkappa \circ \dkappa^{-1})(\mu).
  \end{equation*}
\end{definition}
Note that $\var[Y]=\phi\varf(\mu)$. The support of the members of
\highlight{an} EDF depend only \highlight{on} $\phi\/$ (and not on
$\theta\/$). For a given family, let $C_\phi\/$ be the convex support
of any member of the family with dispersion parameter $\phi\/$. We
define the convex support of the family as
\begin{equation*}
  C_\Phi=\bigcup_{\phi\in\Phi} C_\phi\/.
\end{equation*}

\begin{definition}
  The unit deviance function of an exponential dispersion family  is
  defined as $d: C_\Phi\times \Omega \rightarrow [0,\infty)\/$ with
  \begin{equation}
    \label{eq:unit-deviance}
    d\left(y,\mu\right) = 2 \left[ \sup_{\theta\in\Theta}\{\theta y - \kappa (\theta)\} -
      y \dot{\kappa}^{-1}(\mu) + \kappa\big(\dot{\kappa}^{-1}(\mu)\big)
 \right]\/.
  \end{equation}
\end{definition}

The unit deviance function allows to re-parametrize \eqref{eq:edf-density} as
\begin{equation}
  \label{eq:mean-value-par}
  f(y|\mu,\phi) = c(y,\phi)\exp\left(-\frac{1}{2\phi}d(y,\mu) \right)\/.
\end{equation}
This is known as the mean--value parametrization. When the canonical
space $\Theta\/$ is open, the EDF is said to be regular. In this case
$C_\Phi=\Omega$ and \eqref{eq:unit-deviance} is equivalent to
\begin{equation}
  \label{eq:regular-unit-deviance}
      d\left(y,\mu\right) = 2 \left[ y\{ \dot{\kappa}^{-1}(y)
        -\dot{\kappa}^{-1}(\mu) \}  - 
        \kappa\big(\dot{\kappa}^{-1}(y)\big) +
        \kappa\big(\dot{\kappa}^{-1}(\mu)\big)
 \right]\/.
\end{equation}

\subsection{Weights and Data Aggregation}

In many applications it is useful to include a known positive weight
to each observation. When this is done, the dispersion parameter is
divided by the weight $w$, and \eqref{eq:edf-density} and
\eqref{eq:mean-value-par} become respectively
\begin{align}
  \label{eq:mean-value-par-weights}
f(y|\theta,\phi) &= a(y,\phi/w)\exp\left(\frac{w}{\phi} \left\{y\theta
     - \kappa(\theta) \right\}\right),\quad \mbox{and}\nonumber\\
f(y|\mu,\phi) &= c(y,\phi/w)\exp\left(-\frac{w}{2\phi}d(y,\mu) \right).
\end{align}

There is a useful property of reproductive exponential dispersion
families that allows for data aggregation. J{\o}rgensen's notation (from
\cite{Jorgensen-book}) is very convenient to express this
property: given a fixed exponential family, if \(Y\) has mean \(\mu\)
and density given by \eqref{eq:mean-value-par-weights}, we say that it is
\(ED(\mu,\phi/w)\/\) distributed. The property is then as follows: if
\(Y_1,Y_2,\cdots,Y_n\/\) are independent, and \(Y_i \sim
ED(\mu,\phi/w_i)\/\), then
\begin{equation}
  \label{eq:aggregate-equation}
  \bar{Y}=\frac{w_1Y_1+\cdots+w_nY_n}{w_+}\sim ED(\mu,\phi/w_+),\qquad
  w_+=\sum_{i=1}^n w_i\/.
\end{equation}

\subsection{GLMs}
In a GLM the response variable is assumed to follow an EDF with density
\begin{equation}
  \label{eq:exponential-density}
  f(y|\theta,\phi)=a(y,\phi)\exp\left(
    \frac{w}{\phi} 
    \{y\theta-\kappa( \theta )\}
  \right)\/.
\end{equation}
Note that $\phi\/$ in \eqref{eq:edf-density} corresponds to $\phi/w\/$
in \eqref{eq:exponential-density} which implies that the mean and
variance can be expressed as \(\mu=\kappa'(\theta)\/\) and
\(\sigma^2=\phi \kappa''(\theta)/w\/\), respectively. Here $w\geq 0\/$
is \highlight{known} as the weight. In applications $w\/$ is
\highlight{usually} known and $\phi\/$ needs to be estimated. It is
further assumed that there is a vector of explanatory variables, also
known as covariates, \(\bm{x}=(x_1 \cdots x_p)^T\/\), a vector of
coefficients \(\bm{\beta}=(\beta_0~\beta_1 \cdots \beta_p)^T\/\) and a
function \(g\/\) known as the link function such that
\begin{equation}
  \label{eq:link-equation}
  g(\mu)=\beta_0+x_1\beta_1+\cdots+x_p\beta_p\/.
\end{equation}
It is useful for further developments to express the canonical
parameter \(\theta\/\) in terms of the coefficients. Since $\mu =
\kappa'(\theta) \equiv \dot{\kappa}(\theta)\/$ then:
\begin{align}
  \label{eq:theta-parameters}
  ( g \circ \dkappa ) (\theta) &= \beta_0+x_1\beta_1+\cdots+x_p\beta_p
                              \nonumber \\
  \theta&=( g \circ \dkappa )^{-1}(\beta_0+x_1\beta_1+\cdots+x_p\beta_p)\/.
\end{align}

The population can be divided into different classes according to the
values of the explanatory variables. Thus, given a sample, we can
group together all the observations that share the same values of the
explanatory variables and aggregate them using
\eqref{eq:aggregate-equation}. It is important to mention that with
this grouping there is no loss of information for estimating the mean
since $\bar{Y}\/$ is a sufficient statistic for $\theta\/$ (but not
for $\phi\/$, thus some information is lost for the estimation of
$\phi$). In this sense we say that GLMs are suitable for \emph{data
  aggregation}. \highlight{At the end of Section
  \ref{sec:applied-example}} we illustrate this property with real
data for a unifed GLM.

Possibly after aggregating, let \(m\/\) be the number of classes and
\(\bm{\theta}\in \Theta^m \), where
\( \Theta^m = \left\{\bm{\theta}=(\theta_1 \cdots \theta_m)^T:
  \theta_1,\ldots,\theta_m\in\Theta \right\}\) is the set of all
possible values of the vector $\bm{\theta}$. The density of the sample
can be expressed as
\begin{equation}
  \label{eq:glm-density}
  f(\bm{y} |\bm{\theta},\phi) =
  A(\bm{y},\phi) \exp \left( 
    \frac
    {\bm{y}^TW\bm{\theta} - \bm{1}^T W \bm{\kappa} (\bm{\theta})}
    {\phi}
  \right)\/,\qquad \bm{y}\in\realn^m\/,
\end{equation}
where 
\(\bm{\kappa}(\bm{\theta})=\big(\kappa(\theta_1) \cdots \kappa(\theta_m)\big)^T\/\),
\(W=\mbox{diag}(w_1,\cdots,w_m)\/\), with \(w_i\/\) being the sum of all the 
weights in the \(i\)-th class,
\(\bm{1}=(1 \cdots 1)^T\/\) and \(A(\bm{y},\phi) = \prod_{i=1}^m \big(a(y_i,
\frac{w_i}{\phi})\big)\/\). 

It is useful to reparameterize \eqref{eq:glm-density} in terms of the
mean vector $\bm{\mu}\/$ instead of $\bm{\theta}\/$. Using the mean value
\highlight{parametrization} (this is \eqref{eq:mean-value-par} but substituting
$\phi\/$ for $\phi/w\/$), \eqref{eq:glm-density} can be reparameterized as
\begin{equation}
  \label{eq:glm-deviance-par}
  f(\bm{y}|\bm{\mu},\phi) = C(\bm{y},\phi)\exp \left(-\frac{1}{2\phi}D(\bm{y},\bm{\mu}) \right)\/,
\end{equation}
where $C(\bm{y},\phi)=\prod_{i=1}^mc(y_i,\frac{\phi}{w_i})\/$, and
$D:\Omega^m \times \Omega^m \rightarrow [0,\infty) $ with
\begin{equation}
  \label{eq:glm-deviance-def}
  D(\bm{y},\bm{\mu})=\sum_{i=1}^mw_id(y_i,\mu_i)\/,
\end{equation}
$\Omega^m=\left\{(\mu_1 \cdots \mu_m)^T: \mu_1,\ldots,\mu_m \in \Omega
\right\}$. $D\/$ is called the deviance of the model. Note that
finding the maximum likelihood estimator of $\bm{\beta}$ is equivalent
to finding what value of $\bm{\beta}$ minimizes the deviance. For
further details about the use and properties of the deviance see
\cite{jorgensen-deviance}.


\section{The Unifed Distribution}

The unifed \highlight{family} is the Exponential Dispersion Family
(EDF) generated by the uniform distribution (see Chapters 2 and 3 of
\cite{Jorgensen-book} to see how an EDF can be generated from a moment
generating function). We created the R package \verb|unifed| (see
\cite{unifed-rpackage}) that includes functions to work with the
unifed. In this section we make references to some functions in the
package and we use \verb|this font format| for those references.

To express the density of the unifed \highlight{distribution} we need
the density of the sum of $n$ independent $uniform(0,1)$ random
variables. This corresponds to the Irwin-Hall distribution (see
\cite{continuous-distributions-2}) and its density function is
\begin{equation}
  \label{eq:irwin-hall-density}
  h(y;n) = \frac{1}{(n-1)!}\sum_{k=0}^{\floor{y}} (-1)^k \binom{n}{k}
  (y-k)^{n-1},\qquad y\in[0,n],n\in\nat\correctedComa{.}
\end{equation}

The canonical and index spaces of the unifed family are
$\Theta = \realn$ and $\Phi = \left\{1,\frac{1}{2},\frac{1}{3},\frac{1}{4}\ldots\right\} $, and the cumulant generator is
\begin{equation}
  \label{eq:unifed-cumulant}
  \kappa(\theta)=\left\{
    \begin{array}{ll}
      \log\left(\frac{e^\theta-1}{\theta}\right)& \mbox{if }\theta\neq
                                                  0\\
      0 & \mbox{if }\theta=0
    \end{array}
    \right..
\end{equation}

The density of a unifed \highlight{distribution} with canonical
parameter $\theta$ and dispersion parameter $\phi$ is
\begin{equation}
  \label{eq:unifed-density}
  f(x;\theta,\phi) =
  \frac{h(x/\phi,1/\phi)}{\phi}\exp\left(\frac{x\theta -
      \kappa(\theta)}{\phi}\right),
\end{equation}
where $h$ and $\kappa$ are as in \eqref{eq:irwin-hall-density} and
\eqref{eq:unifed-cumulant}, respectively and
$x\in[0,1],\theta\in\realn,
\phi\in\left\{1,\frac{1}{2},\frac{1}{3},\ldots\right\}$. We denote the
unifed distribution with canonical parameter $\theta$ and dispersion
parameter $\phi$ with $unifed(\theta,\phi)$.

\begin{figure}[h]
  \centering
  \begin{subfigure}[b]{0.495\textwidth}
    \includegraphics[width=\textwidth]{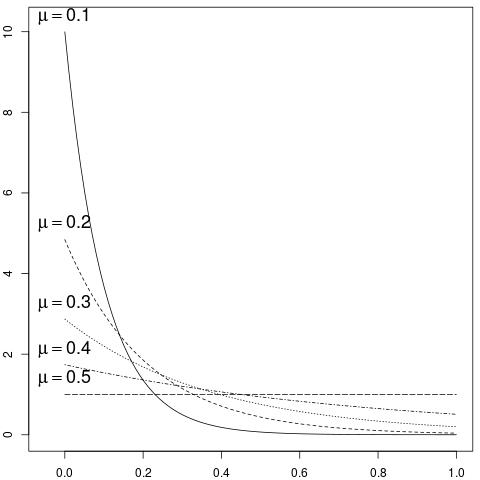}
  \end{subfigure}
  \begin{subfigure}[b]{0.495\textwidth}
    \includegraphics[width=\textwidth]{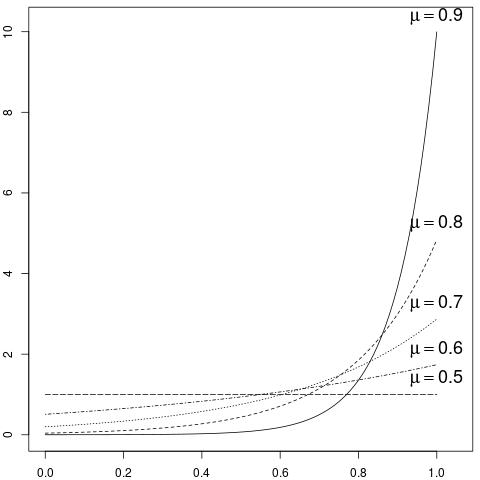}
  \end{subfigure}
  \caption{Density of the unifed for different values of its mean $\mu$}
  \label{fig:unifed-density-plot}
\end{figure}

The unifed package does not contain an implementation of
\eqref{eq:unifed-density}. This is because we did not find a
numerically stable way to compute $h$. To show this, the package
includes the function \verb|dirwin.hall| that computes $h$. Table
\ref{tab:irwin-hall-overflow} shows the results we get by calling this
function with $n$ set to 50 and varying the values of $y$. The changes
of sign indicate that a float overflow is happening.

\begin{table}[h]
\centering
\begin{tabular}{|c|c|}
  Code & Result\\
  \hline
  \verb|dirwin.hall(35,50)| & 0.0674864\\
  \hline
  \verb|dirwin.hall(36,50)| & -13.12745 \\
  \hline
  \verb|dirwin.hall(37,50)| & 45.44388 \\
  \hline
  \verb|dirwin.hall(38,50)| & -37.44488 \\
  \hline
\end{tabular}
\caption{Float overflow of the Irwin-Hall implementation.}
\label{tab:irwin-hall-overflow}
\end{table}

\highlight{The} package calls unifed distribution the \highlight{one-parameter} special
case of \eqref{eq:unifed-density} where $\phi=1$, which we denote with
$unifed(\theta)$. This simplifies the density to

\begin{equation}
  \label{eq:unifed-density-2}
  f(x;\theta) = \left\{
  \begin{array}{ll}
    \frac{\theta}{e^\theta - 1} e^{x \theta} & \mbox{if } \theta \neq 0\\
    1 & \mbox{if } \theta = 0
  \end{array}
  \right. \quad \mbox{for } x \in (0,1).
\end{equation}

The functions \verb|dunifed|, \verb|punifed|, \verb|qunifed| and
\verb|runifed|, give the density, distribution, quantile and
simulation functions, respectively of this simplified version. The mean
and variance of each element of the family are given by

\begin{align}
  \label{eq:unifed-mean-var}
  \esp[X]&= \dkappa(\theta) = \left\{
  \begin{array}{ll}
   \frac{(\theta-1)e^\theta + 1}{\theta(e^\theta -1)} & \mbox{if
                                                        }\theta \neq
                                                        0\\
    \frac{1}{2} & \mbox{if }\theta=0
  \end{array}
  \right., \qquad \\
  \var[X]&= \ddkappa(\theta)=\left\{
    \begin{array}{ll}
      \left(\frac{ e^{2\theta} - (\theta+2)e^\theta + 1}
      {\theta^2 (e^\theta-1)^2}\right)& \mbox{if }\theta \neq 0 \\
      \frac{1}{12} & \mbox{if }\theta=0
    \end{array}
    \right.,
\end{align}
where $\dkappa$ and $\ddkappa$ are the first and second derivative of
$\kappa$, respectively. We have not been able to find an analytical
expression for the inverse function $\dot{\kappa}^{-1}$. Thus, it
has not been possible either to find analytical expressions for the
variance function and unit deviance of the unifed. Nevertheless, the
\emph{unifed} package contains the function
\verb|unifed.kappa.prime.inverse| that uses the Newthon Raphson method
to implement the inverse of $\dkappa$. This allows us to get a
numerical solution for the variance function by using the relation
$\varf(\mu) = \ddot{\kappa} (\dot{\kappa}^{-1}(\mu) )$. This is
implemented in the function \verb|unifed.varf|.

Similarly, since the unifed is a regular EDF (see Chapter 2 of
\cite{Jorgensen-book}), we can compute the unit deviance by using the
relation
\begin{equation}
  \label{eq:regular-unit-deviance-1}
  d(y,\mu)=2\left[y\{\dot{\kappa}^{-1}(y)-\dot{\kappa}^{-1}(\mu)\}-\kappa(\dot{\kappa}^{-1}(y))+\kappa(\dot{\kappa}^{-1}(\mu))\right].
\end{equation}
The function \verb|unifed.unit.deviance| computes the unit deviance
using \eqref{eq:regular-unit-deviance-1}. As mentioned in Section
\ref{sec:edf-glms}, the unit deviance can be used to reparametrize the
distribution in terms of it's mean and dispersion parameter. We denote
with $unifed^*(\mu,\phi)$ the unifed distribution with mean $\mu$ and
dispersion parameter $\phi$ and when $\phi=1$, we write simply
$unifed^*(\mu)$.

Figure \ref{fig:unifed-density-plot} shows plots of the unifed
distribution for different values of its mean. We can see that except
for $\mu=0.5$, it is always monotone. For $\mu < 0.5$ it is strictly
decreasing and the mode is at zero. For $\mu > 0.5$ it is strictly
increasing and the mode is at one. The R code used for producing this
plot can be found in \cite{snippet-unifed-density-plot}.

\subsection{Maximum Likelihood Estimation}

Suppose you have an independent and identically distributed sample
$X_1,\ldots,X_n$ coming from a $unifed(\theta)$ distribution and you
want to compute the maximum likelihood estimator (mle) $\hat{\theta}$
of $\theta$. The derivative of the log-likelihood function is given by
\begin{align*}
  \ell'(\theta|X_1,\ldots,X_n)&= n \frac{(1-\theta)e^\theta -
                               1}{\theta(e^\theta - 1)} + \sum_{i=1}^n X_i \\
  &= - n \dkappa(\theta) + \sum_{i=1}^n X_i.
\end{align*}
Making the expression above equal to zero and solving for \highlight{$\theta$,}
the mle for $\theta$ is given by
\begin{equation}
  \label{eq:unifed-mle}
  \hat{\theta} = \dkappa^{-1}\left( \bar{X}\right),
\end{equation}
where $\bar{X}=\sum_{i=1}^nX_i / n$. The function \verb|unifed.mle| in
the unifed R package computes the mle using \eqref{eq:unifed-mle}. It
is possible to use the unifed distribution as the response
distribution of a GLM. In this case, $\phi$ must be fixed to one and
the weight of each class is the number of observations in the
class. The mle \highlight{$\hat{\beta}$ of the regression
  coefficients} can be found using iterative weighted least
squares. In Section 2.5 of \cite{macnelbook}, they show that this
method works for any response distribution whose density can be
expressed as \highlight{\eqref{eq:exponential-density}}. Thus, the method also
works for the unifed. The unifed R package (\cite{unifed-rpackage})
provides the function \verb|unifed| that returns a family object than
can be used inside the \verb|glm| function.

\section{An Illustrative Example}
\label{sec:applied-example}
In this section we apply a unifed GLM to a publicly available dataset.
The data appears in \cite{glm-insurance-book}. It is based on 67,856
one--year auto insurance policies from 2004 or 2005. \highlight{The
  dataset} can be downloaded from the companion site of the book (see
\cite{glm-insurance-book}). Table
\ref{tab:vehicle-insurance-description} shows the description of the
variables as provided at the website.

{\linespread{1}
\small
\begin{table}[t]
  \centering
  \begin{tabular}{|ll|}
    \hline
    Variable name & Description\\
    \hline
    \hline
    \verb+veh_value+&	vehicle value, in \$10,000s \\
    \hline
    \verb+exposure+ & 0-1 \\
    \hline
    \verb+clm+ & occurrence of claim (0 = no, 1 = yes)\\
    \hline
    \verb+numclaims+ & number of claims\\
    \hline
    \verb+claimcst0+ & 	claim amount (0 if no claim)\\
    \hline
    \verb+veh_body+ & vehicle body, coded as\\				
                    & BUS\\
                    & CONVT = convertible  \\
                    & COUPE   \\
                    & HBACK = hatchback                  \\
                    & HDTOP = hardtop\\
                    & MCARA = motorized caravan\\
                    & MIBUS = minibus\\
                    & PANVN = panel van\\
                    & RDSTR = roadster\\
                    & SEDAN    \\
                    & STNWG = station wagon\\
                    & TRUCK           \\
                    & UTE - utility\\
    \hline
    \verb+veh_age+ & age of vehicle: 1 (youngest), 2, 3, 4\\
    \hline
    \verb+gender+ & gender of driver: M, F \\
    \hline
    \verb+area+ & driver's area of residence: A, B, C, D, E, F \\
    \hline
    \verb+agecat+ & driver's age category: 1 (youngest), 2, 3, 4, 5, 6\\
    \hline
  \end{tabular}
  \caption{Vehicle insurance variables}
  \label{tab:vehicle-insurance-description}
\end{table} }

We are interested in modeling the exposure; which is the proportion of
time of the year in which the insurance policy is in-force for a given
client. We use \verb|gender|, \verb|agecat|, \verb|area| and
\verb|veh_age| as the explanatory variables.

The R code used to obtain the results that follow can be found in
\cite{snippet-vehicle-example}.

The data was aggregated using \eqref{eq:aggregate-equation} and a
unifed GLM was fit to it. Table \ref{tab:applied-example-glm-summary}
(exported from R using the package \verb|xtable| \cite{xtable}) shows the
summary provided by the \verb|glm| function of R. We see that all the
variables included have at least one significant class.

A $\chi^2$ test for goodness of fit is commonly used for GLMs. The
null hypothesis is that the data is distributed according to the
fitted GLM. Assuming the null hypothesis for this example implies that
the residual deviance reported at the bottom of Table
\ref{tab:applied-example-glm-summary} follows a $\chi^2$ distribution
with 273 degrees of freedom. The p-value for this example is
$\mathbb{P}(\chi_{273}^2\ge 297.86)=0.14$. Now, the detail with this
test is that the $\chi^2$ distribution for the residual deviance is
asymptotic on the smallest weight of all classes going to infinity
(see \cite[Section 3.6]{jorgensen-deviance}). The smallest observed
weight here is 4 and it corresponds to the class with \verb|gender|=F,
\verb|agecat|=6, \verb|area|=F and \verb|veh_age|=1. Therefore the
$\chi^2$ test for this example is not reliable.

\begin{figure}[h]
  \centering
  \includegraphics[width=0.65\textwidth]{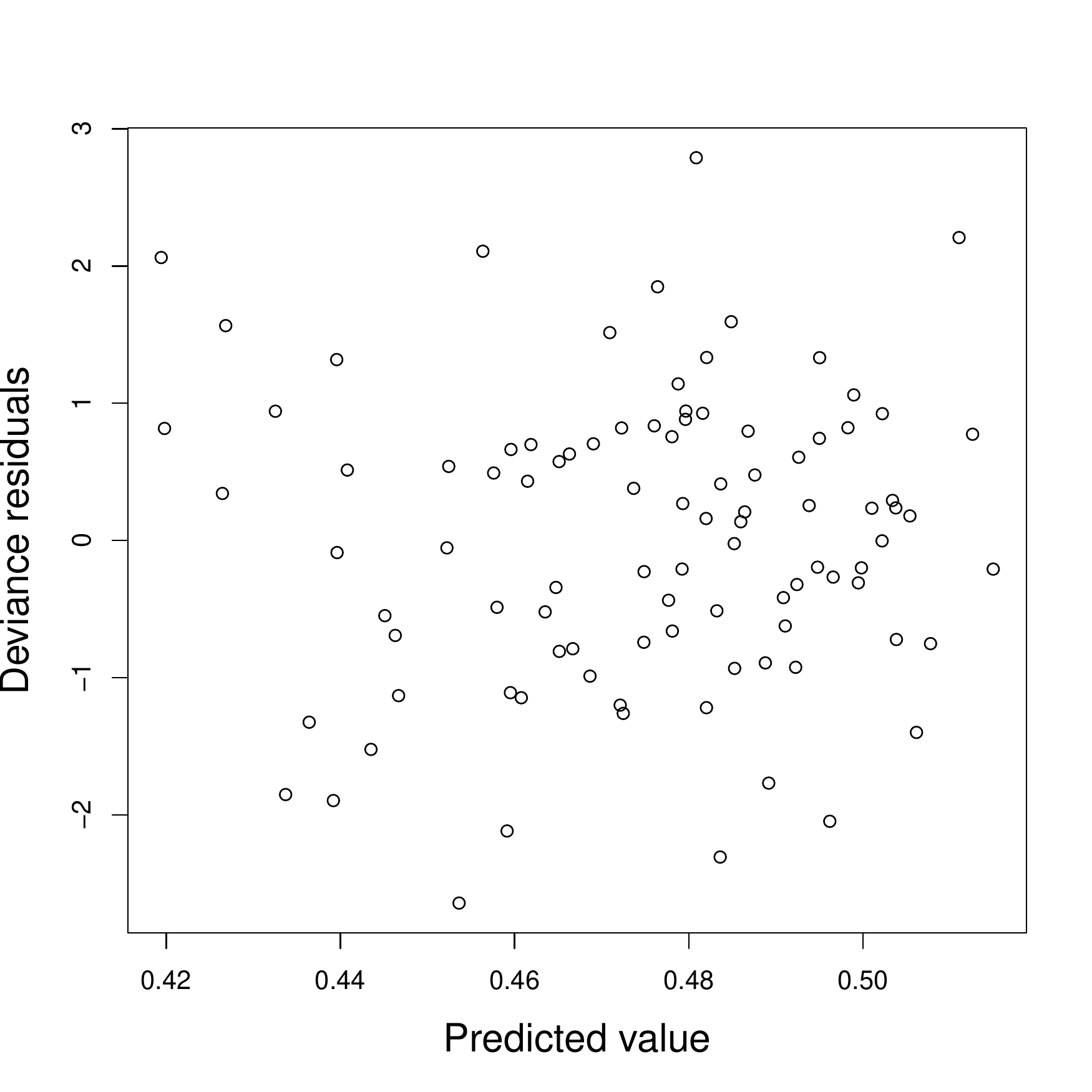}
  \caption{Residuals of Unifed Regression}
  \label{fig:applied-example-residuals}
\end{figure}

Figure \ref{fig:applied-example-residuals} shows the deviance
residuals of this model. It suggests a good fit since they do not show
any apparent pattern.

\begin{table}[ht]
\centering
\begin{tabular}{rrrrr}
  \hline
 & Estimate & Std. Error & z value & Pr($>$$|$z$|$) \\ 
  \hline
(Intercept) & -0.3319 & 0.0197 & -16.84 & 0.0000 *** \\ 
  genderM & 0.0288 & 0.0090 & 3.20 & 0.0014 ** \\ 
  agecat2 & 0.0011 & 0.0184 & 0.06 & 0.9518 \\ 
  agecat3 & 0.0530 & 0.0178 & 2.97 & 0.0029 **\\ 
  agecat4 & 0.0583 & 0.0178 & 3.28 & 0.0010 **\\ 
  agecat5 & 0.1042 & 0.0189 & 5.51 & 0.0000 ***\\ 
  agecat6 & 0.0692 & 0.0210 & 3.30 & 0.0010 ***\\ 
  areaB & 0.0239 & 0.0135 & 1.77 & 0.0761 .\\ 
  areaC & 0.0014 & 0.0121 & 0.11 & 0.9086 \\ 
  areaD & 0.0053 & 0.0157 & 0.34 & 0.7337 \\ 
  areaE & 0.0120 & 0.0175 & 0.68 & 0.4948 \\ 
  areaF & 0.0879 & 0.0214 & 4.10 & 0.0000 ***\\ 
  veh\_age2 & 0.1708 & 0.0138 & 12.40 & 0.0000 ***\\ 
  veh\_age3 & 0.1613 & 0.0133 & 12.16 & 0.0000 ***\\ 
  veh\_age4 & 0.1549 & 0.0134 & 11.53 & 0.0000 ***\\ 
  \hline
    \multicolumn{5}{l}{Signif. codes:  0 ‘***’ 0.001 ‘**’ 0.01 ‘*’ 0.05 ‘.’ 0.1 ‘ ’ 1}\\
  \multicolumn{5}{c}{(Dispersion parameter for unifed family taken to be 1)}\\
  Null deviance: & \multicolumn{4}{l}{585.47  on 287  degrees of freedom }\\
  Residual deviance: & \multicolumn{4}{l}{297.86  on 273  degrees of freedom}\\
  \hline
\end{tabular}
\caption{Summary of Unifed GLM}
\label{tab:applied-example-glm-summary}
\end{table}

\subsection*{\highlight{Verifying Data Aggregation:}}

\highlight{We} now fit the same model \highlight{as} in the previous
section but without aggregating the data. Table
\ref{tab:applied-example-glm-summary-no-aggregation} shows the summary
of the model from R. The code used to generate this table can be found
in \cite{snippet-vehicle-example}.

By comparing Tables \ref{tab:applied-example-glm-summary} and
\ref{tab:applied-example-glm-summary-no-aggregation} one can see that
the estimated coefficients are the same in both cases. Thus, even
though the deviance of both models differ, they give the same mle for
the coefficients. This shows what we mean with data aggregation.

\begin{table}[ht]
\centering
\begin{tabular}{rrrrr}
  \hline
 & Estimate & Std. Error & z value & Pr($>$$|$z$|$) \\ 
  \hline
(Intercept) & -0.3319 & 0.0197 & -16.84 & 0.0000 ***\\ 
  genderM & 0.0288 & 0.0090 & 3.20 & 0.0014 **\\ 
  agecat2 & 0.0011 & 0.0184 & 0.06 & 0.9518 \\ 
  agecat3 & 0.0530 & 0.0178 & 2.97 & 0.0029 **\\ 
  agecat4 & 0.0583 & 0.0178 & 3.28 & 0.0010 **\\ 
  agecat5 & 0.1042 & 0.0189 & 5.51 & 0.0000 ***\\ 
  agecat6 & 0.0692 & 0.0210 & 3.30 & 0.0010 ***\\ 
  areaB & 0.0239 & 0.0135 & 1.77 & 0.0761 .\\ 
  areaC & 0.0014 & 0.0121 & 0.11 & 0.9086 \\ 
  areaD & 0.0053 & 0.0157 & 0.34 & 0.7337 \\ 
  areaE & 0.0120 & 0.0175 & 0.68 & 0.4948 \\ 
  areaF & 0.0879 & 0.0214 & 4.10 & 0.0000 ***\\ 
  veh\_age2 & 0.1708 & 0.0138 & 12.40 & 0.0000 ***\\ 
  veh\_age3 & 0.1613 & 0.0133 & 12.16 & 0.0000 ***\\ 
  veh\_age4 & 0.1549 & 0.0134 & 11.53 & 0.0000 ***\\ 
  \hline
  \multicolumn{5}{l}{Signif. codes:  0 ‘***’ 0.001 ‘**’ 0.01 ‘*’ 0.05 ‘.’ 0.1 ‘ ’ 1}\\
  \multicolumn{5}{c}{(Dispersion parameter for unifed family taken to be 1)}\\
  Null deviance: & \multicolumn{4}{l}{113445  on 67855  degrees of freedom}\\
  Residual deviance: & \multicolumn{4}{l}{113158  on 67841  degrees of freedom}\\
  \hline
\end{tabular}
\caption{Summary of Unifed GLM without Data Aggregation}
\label{tab:applied-example-glm-summary-no-aggregation}
\end{table}

\section{Comparison Between the Unifed GLM and the Beta Regression}

The beta regression (\cite{beta-regression}) is a versatile model for
applications with a response variable on the unit interval. Moreover,
the well documented R package \verb|betareg|
(\cite{r-beta-regression}) makes it a practical tool in many
applications.

\subsection{The Beta Regression}

The density of the beta distribution contains a large variety of
shapes. In \cite{beta-regression} the beta density is reparameterized
as

\begin{equation}
  \label{eq:beta-density}
  f(y) = \frac{\Gamma(\phi)}{\Gamma(\mu \phi) \Gamma( ( 1 - \mu ) \phi ) }
  y^{\mu \phi -1} ( 1 - y )^{ (1-\mu)\phi-1 },\qquad 0 < y < 1,
\end{equation}
with $0<\mu<1$ and $\phi > 0$, and the distribution is denoted by
$\Beta(\mu,\phi)$. Under this parametrization, if $Y \sim
\Beta(\mu,\phi)$, the mean and variance are
\begin{equation}
  \label{eq:beta-mean-variance}
  \esp[Y] = \mu \quad \mbox{and} \quad \var[Y]=\frac{\mu (1-\mu)}{1+\phi}.
\end{equation}

Here $\phi$ is called the precision parameter of the distribution. In the
beta regression model it is assumed that the response variable is a
vector $Y=(Y_1,\ldots,Y_m)$, in which $Y_i\sim \Beta(\mu_i,\phi)$ for
$i=1,\ldots,m$. The $Y_i's$ are assumed independent to each other. The
explanatory variables are incorporated to the model through the
relation
\begin{equation*}
  g(\mu_i) = \bm{x_i}^T\bm{\beta},
\end{equation*}

where $\bm{\beta}$  is a vector of parameters and $\bm{x_i}$ is a vector
of \highlight{regressors}. $g:(0,1)\rightarrow \realn$ is invertible and is
called the link function.

Then \cite{beta-double-regression} generalized this model to allow the
precision parameter to vary among classes in a similar way to the
double generalized linear models (see \cite{dglms}). More
specifically, in this case the response vector $Y=(Y_1,\ldots,Y_m)$ is
such that $Y_i\sim \Beta(\mu_i,\phi_i)$, independently and
\begin{align*}
  g_1(\mu_i) &= \bm{x_i}^T\bm{\beta}, \\
  g_2(\phi_i) &= \bm{z_i}^T\bm{\gamma}, 
\end{align*}
where $\bm{\beta}$ and $\bm{\gamma}$ are regression coefficients.

These regression models offer great flexibility when the response
variable lies in the interval $(0,1)$, and both are implemented in the
R package \verb|betareg| (\cite{r-language},
\cite{r-beta-regression}).

The beta distribution is not an \highlight{EDF} and therefore the beta
regression is not a GLM. Nevertheless \highlight{the} parametrization chosen by
the authors of the model along with \eqref{eq:beta-mean-variance} give
it a similar look and feel.

\subsection{On the Difficulties of Data Aggregation for the Beta Regression}

Data aggregation gives a practical advantage when working with large
datasets. For GLMs this is straightforward due to two properties of
$\bar{Y}$ in \eqref{eq:aggregate-equation}:

\begin{itemize}
\item $\bar{Y}$ is a sufficient statistic for $\mu$
\item The distribution of $\bar{Y}$ belongs to the same family
  \highlight{as} the $Y_i$'s in \eqref{eq:aggregate-equation}.
\end{itemize}

We do not know any statistic with these two properties for the beta
distribution. For instance, let $Y_1,\ldots,Y_n$ be an i.i.d sample
from a $\Beta(\mu,\phi)$ distribution. The joint likelihood function
of this sample is then

\begin{equation*}
  f(\bm{y}) = \left(\frac{\Gamma(\phi)}{\Gamma(\mu \phi) \Gamma( ( 1 - \mu ) \phi ) }\right)^n
  \left(\prod_{i=1}^ny_i\right)^{\mu \phi -1} \left(\prod_{i=1}^n(1 - y_i)\right)^{ (1-\mu)\phi-1 }, 
\end{equation*}

where $\bm{y}=(y_1,\ldots,y_n)$. This density can be rearranged as
follows
\begin{equation*}
  f(\bm{y}) = \left(\frac{\Gamma(\phi)}{\Gamma(\mu \phi) \Gamma( ( 1 - \mu ) \phi ) }\right)^n
  \left[\prod_{i=1}^n\frac{(1-y_i)^{\phi-1}}{ y_i}\right]
  \left(\prod_{i=1}^n \frac{y_i}{1-y_i} \right)^{\mu\phi}
\end{equation*}

The factorization theorem (see \cite[Chapter 7]{intro-math-stats}), implies that
$T=\prod_{i=1}^n \frac{y_i}{1-y_i}$ is sufficient for $\mu$. Now, the
distribution of $T$, which is not beta, would be needed to use $T$ for
data aggregation. In other words, a regression model whose response
distribution is a family that includes the distribution of $T$ for
every $n$ would need to be developed.

\subsection{Differences Between the Unifed GLM and the Beta Regression}

\begin{figure}[h]
  \centering
  \includegraphics[width=0.65\textwidth]{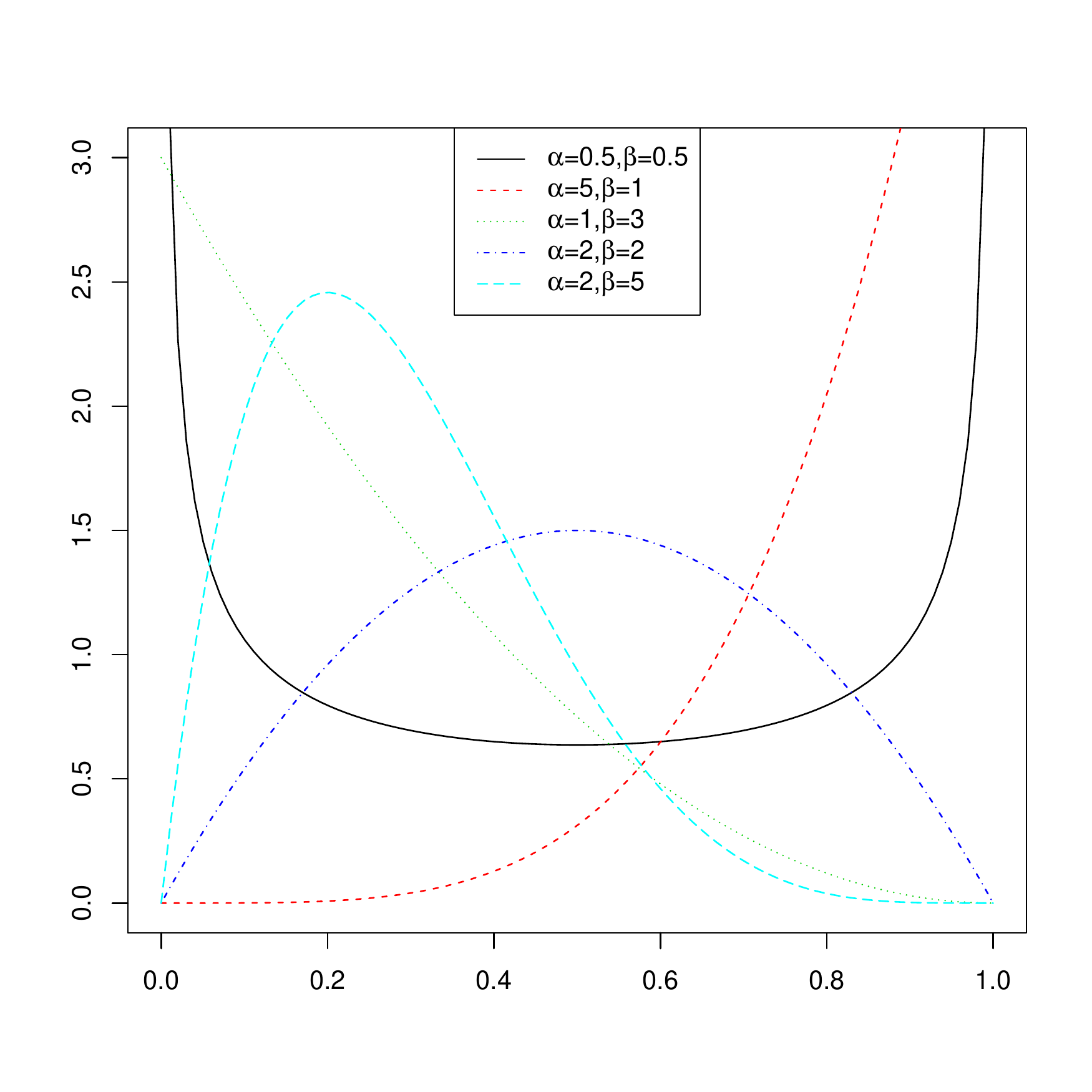}
  \caption{Beta Densities for Different Values of $\alpha$ and $\beta$}
  \label{fig:beta-densities}
\end{figure}

The unifed density does not have the variety of shapes that the beta
density has. To see this\correctedComa{,} compare the shapes shown in Figure
\ref{fig:unifed-density-plot} with the shapes for the beta
distribution shown in Figure \ref{fig:beta-densities}
\cite{snippet-beta-density-plot}. Thus, the beta regression is able to
adapt to more shapes than a unifed GLM and even more so if \highlight{regressors}
are used for the dispersion parameter.

In those cases where a beta regression and a unifed GLM give similar
good fit, the parsimony principle suggests to pick the unifed GLM\correctedComa{ }
since it has one parameter less; the dispersion parameter is
known for the unifed GLM.

From a numerical point of view, the unifed GLM has the advantage that
it is possible to use \eqref{eq:aggregate-equation} for data
reduction. This is a practical advantage when dealing with large
datasets specially if simulations of the response vector need to be
performed.
\section{Conclusion}

This paper introduced a new distribution called unifed. It is the
Exponential Dispersion Family generated by the uniform
distribution. It allows to fit a GLM for responses on the unit
interval (0,1). An R package for working with this distribution is
provided.

We made a comparison to the beta regression, which is another
regression model for responses on the unit interval. It provides more
flexible shapes and therefore it can give better fit than a unifed GLM
in many situations. In contrast, the unifed GLM is suitable for data
aggregation which is a practical advantage when working with large
datasets.

An application using publicly available data was presented.  

\section*{Abbreviations}
\begin{description}
\item[EDF:] Exponential Dispersion Family
\item[GLM:] Generalized Linear Model
\item[mle:] Maximum Likelihood Estimator
  
\end{description}

\section*{Declarations}
\subsection*{Availability of data and material} The data used for
the example in this article is publicly available and it can be
downloaded from
\url{www.businessandeconomics.mq.edu.au/our_departments/Applied_Finance_and_Actuarial_Studies/acst_docs/glms_for_insurance_data/data/car.csv}.
\subsection*{Competing interests}
The author declares that they have no competing interests.

\subsection*{Funding} 
Not applicable.
\subsection*{Authors' contributions}
All contributions were made by the author of the article, Oscar Alberto Quijano Xacur.
\subsection*{Acknowledgements}
Not applicable.
\clearpage
\bibliographystyle{plainnat}
\bibliography{referencias}

\end{document}